\documentclass[aps,floatfix,preprint,showpacs,preprintnumbers,nofootinbib,superscriptaddress,natbib]{revtex4}

\usepackage{graphicx,float}
\usepackage{epsfig}		
\usepackage{dcolumn}
\usepackage{bm}
\usepackage{subfigure}

\usepackage[all]{xy}
\usepackage{amsmath,upgreek}
\usepackage{amssymb}

\usepackage{pdfpages}
\usepackage{color,soul}
\usepackage{graphicx,epstopdf}
\usepackage[colorlinks=true,
 linkcolor=red,
 urlcolor=purple,
 citecolor=blue,hyperindex]{hyperref}
\def\0{\mbox{\tiny $0$}}
\def\1{\mbox{\tiny $1$}}
\def\2{\mbox{\tiny $2$}}
\def\3{\mbox{\tiny $3$}}
\def\4{\mbox{\tiny $4$}}
\def\5{\mbox{\tiny $5$}}
\def\6{\mbox{\tiny $6$}}
\def\7{\mbox{\tiny $7$}}
\def\8{\mbox{\tiny $8$}}
\def\9{\mbox{\tiny $9$}}

\def\f14{\mbox{\tiny $\frac{1}{4}$}}

\def\bb#1{\mbox{\footnotesize $(#1)$}}

\begin{document}

\title{Non-classicality from the phase-space flow analysis of the Weyl-Wigner quantum mechanics}

\author{Alex E. Bernardini}
\email{alexeb@ufscar.br}
\altaffiliation[On leave of absence from]{~Departamento de F\'{\i}sica, Universidade Federal de S\~ao Carlos, PO Box 676, 13565-905, S\~ao Carlos, SP, Brasil.\\}
\author{Orfeu Bertolami}
\email{orfeu.bertolami@fc.up.pt}
\altaffiliation[Also at~]{Centro de F\'isica do Porto, Rua do Campo Alegre, 4169-007, Porto, Portugal.} 

\affiliation{Departamento de F\'isica e Astronomia, Faculdade de Ci\^{e}ncias da
Universidade do Porto, Rua do Campo Alegre, 4169-007, Porto, Portugal.}
\vspace{1 cm}
\date{\today}

\begin{abstract}
A fluid analog of the information flux in the phase-space associated to purity and von Neumann entropy are identified in the Weyl-Wigner formalism of quantum mechanics.
Once constrained by symmetry and positiveness, the encountered continuity equations provide novel quantifiers for non-classicality (non-Liouvillian fluidity) given in terms of quantum decoherence, purity and von Neumann entropy fluxes.
Through definitions in the Weyl-Wigner formalism, one can identify the quantum fluctuations that distort the classical-quantum coincidence regime, and the corresponding quantum information profile, whenever some bounded $x-p$ volume of the phase-space is specified.
The dynamics of anharmonic systems is investigated in order to illustrate such a novel paradigm for describing quantumness and classicality through the flux of quantum information in the phase-space.
\end{abstract}
\vspace{1 cm}

\pacs{03.65.-w, 03.65.Sq}
\keywords{quantum to classical transitions - phase-space - Wigner representation - quantum information}
\date{\today}
\maketitle
\renewcommand{\baselinestretch}{1}

Decoherence and entropy play a crucial role in the understanding of the frontiers between classical and quantum descriptions of Nature in quantum mechanics (QM) \cite{02A}.
These concepts are also at the heart of quantum statistical properties that must be considered in a theory of quantum measurement \cite{01A,02A,03A}.
In this context, the Weyl-Wigner (WW) formalism \cite{Wigner} is particularly relevant.
Indeed, the WW phase-space representation of QM is akin to the formalism of statistical mechanics.
Pragmatically, it establishes a complementary phase-space formulation of QM that provides a straightforward access to quantum information issues without modifying its predictive power in terms of quantum observables and expectation values.

In the context of the WW formalism, the Weyl transform of a quantum operator, $\hat{O}$, is defined as
\begin{eqnarray}
O^W(x, p)
&=& \hspace{-.2cm} \int \hspace{-.15cm}dy\,\exp{\left[2\,i \,p\, y/\hbar\right]}\,\langle x - y | \hat{O} | x + y \rangle=\hspace{-.2cm} \int \hspace{-.15cm} dy \,\exp{\left[-2\, i \,x\, y/\hbar\right]}\,\langle p - y | \hat{O} | p + y\rangle,
\end{eqnarray}
and the Wigner function, $W(x, p)$, is identified as the Weyl transform of a density matrix $\hat{\rho} = |\psi \rangle \langle \psi |$, given by
\begin{equation}
 h^{-1} \hat{\rho} \to  W(x, p) =  (\pi\hbar)^{-1} 
\int \hspace{-.15cm}dy\,\exp{\left[2\, i \, p \,y/\hbar\right]}\,
\psi(x - y)\,\psi^{\ast}(x + y).
\end{equation}

The averaged values associated to $\hat{O}$ can be computed in terms of the trace of the product of the two operators, $\hat{\rho}$ and $\hat{O}$, identified by the phase-space integral of the product of their Weyl tranforms \cite{Wigner,Case}\footnote{That is generically expressed by
\begin{equation}
Tr_{\{x,p\}}\left[\hat{O}_1\hat{O}_2\right] =  
\int \hspace{-.15cm}\int \hspace{-.15cm} {dx\,dp} \,O^W_1(x, p)\,O^W_2(x, p),
\end{equation}
for any product of two operators, $\hat{O}_{1}$ and $\hat{O}_{2}$.}
\begin{equation}
Tr_{\{x,p\}}\left[\hat{\rho}\hat{O}\right] \to \langle O \rangle = 
\int \hspace{-.15cm}\int \hspace{-.15cm} {dx\,dp}\,W(x, p)\,{O^W}(x, p),
\label{five}
\end{equation}
which sets the generalization of such quasi-probability distributions, where $Tr[\hat{\rho}]=1$.
The above operational properties provide, for instance, an expression for the quantum purity, $\mathcal{P}$, defined as
\begin{equation}
Tr_{\{x,p\}}[\hat{\rho}^2] \to  \mathcal{P} = 2\pi\int \hspace{-.15cm}\int \hspace{-.15cm} {dx\,dp}\,W^2(x, p),
\label{pureza}
\end{equation}
where the factor $2\pi$ appears in order to satisfy the density matrix condition, $Tr[\hat{\rho}^2] = 1$, for pure states, and $Tr[\hat{\rho}^2] < 1$, for statistical mixtures.
{In fact, the theoretical framework of the QM in the phase-space is more completely addressed if one also accounts for the Moyal's picture of QM} \cite{Moyal}, {which exhibits the operational noncommutativity between coordinate and momentum through the Moyal {\em star}-product so to recover the WW formalism.}

Of course, a phase-space formulation of QM is not necessarily and exclusively described by Wigner functions \cite{Ballentine}.
Indeed, given that the Wigner function description admits negative amplitude values, it cannot be strictly interpreted as probability distributions, and thus alternative phase-space distribution functions have been considered \cite{Husimi,Glauber,Sudarshan,Carmichael,Callaway} so to ensure a corresponding non-negative probability interpretation.
{For instance, in the optical tomographic probability representation of QM} \cite{Amosov}, {the Radon transform} \cite{Radon} {of the Weyl-Wigner-Moyal equation is always positive, even for Wigner functions assuming negative values.
In particular, the associated symplectic tomographyc probability form of the Weyl-Wigner-Moyal equation works as a classical approach to quantum systems} \cite{Mancini}, {which is highly representative in the context of entropy and information dynamics.} 

Turning back to the fluid-analog framework, our analysis is particularly concerned to the quantum aspects of physical systems \cite{Steuernagel3,Ferraro11} that can be described by the time evolution of the Wigner function, $W(x,\,p;\,\tau)$, when cast in the form of a vector flux $\mathbf{J}(x,\,p;\,\tau)$ \cite{Donoso12,Domcke,Waldron}, that drives the flow of $W(x,\,p;\,\tau)$ in the phase-space.
For the flow field identified by the phase-space component directions, $\mathbf{J} = J_x\,\hat{x} + J_p\,\hat{p}$, where $\hat{p} = \hat{p}_x$, the equivalent of the Schr\"{o}dinger equation in phase-space can be written in terms of a continuity equation \cite{Case,Ballentine,Steuernagel3}:
\begin{equation}
\frac{\partial W}{\partial \tau} + \frac{\partial J_x}{\partial x}+\frac{\partial J_p}{\partial p} \equiv
\frac{\partial W}{\partial \tau} + \mbox{\boldmath $\nabla$}\cdot \mathbf{J} =0,
\label{zeqnz51}
\end{equation}
the so-called quantum Liouville equation, where
\begin{equation}
J_x(x,\,p;\,\tau)= \frac{p}{m}\,W(x,\,p;\,\tau) \quad\mbox{and}\quad 
J_p(x,\,p;\,\tau) =-\frac{\partial U(x)}{\partial x}W(x,\,p;\,\tau) + \Delta J_p(x,\,p;\,\tau),
\label{zeqnz500BB}
\end{equation}
so that $U(x)$ is the potential and 
\begin{equation}
\Delta J_p(x,\,p;\,\tau)= -\sum_{k=1}^{\infty} \left(\frac{i\,\hbar}{2}\right)^{2k}\frac{1}{(2k+1)!} \, \left(\frac{\partial~}{\partial x}\right)^{2k+1}\hspace{-.5cm}U(x)\,\left(\frac{\partial~}{\partial p}\right)^{2k}\hspace{-.3cm}W(x,\,p;\,\tau)
\label{zeqnz500}
\end{equation}
depicts the distortion due to the quantum features.
From the evaluation of coordinate and momentum integrations, the marginals of the Wigner function yield, respectively, the probability density, $\vert \psi(x;\,\tau)\vert^2$, and the momentum distribution, $\vert \varphi(p;\,\tau)\vert^2$,
where
\begin{equation}
\varphi(p;\,\tau) = \int_{-\infty}^{+\infty}\hspace{-.2 cm}{dx}\,\psi(x;\,\tau)\,\exp(i\,p\,x).
\label{zeqnz52}
\end{equation}
Considering that the evolution of the probability density and the momentum distribution leads to the marginal continuity equations,
\begin{equation}
\frac{d}{d\tau} \vert \psi(x;\,\tau)\vert^2 = \int_{-\infty}^{+\infty}\hspace{-.2 cm}{dp}\,\partial_x J_x(x,\,p;\,\tau) = \partial_x j_x(x;\,\tau),
\label{zeqnz53A}
\end{equation}
\begin{equation}
\frac{d}{d\tau} \vert\varphi(p;\,\tau)\vert^2 = \int_{-\infty}^{+\infty}\hspace{-.2 cm}{dx}\,\partial_p J_p(x,\,p;\,\tau) = \partial_p j_p(p;\,\tau),
\label{zeqnz53}
\end{equation}
fluid dynamics analogs related to non-classicality, entropy and purity fluxes, can be constructed\footnote{Notice that the lower order moments of the Wigner function can also be helpful in a subliminal fluid dynamics analogy.
The fluid particle density $\rho_{_\psi} \equiv \vert \psi(x;\,\tau)\vert^2$ and the fluid averaged velocity (normalized current density) $v^{\psi}_x(x;\,\tau) \equiv {j_x(x;\,\tau)}/{\rho_{\psi}(x;\,\tau)}$ can be respectively identified by 
\begin{equation}
\int_{-\infty}^{+\infty}\hspace{-.2 cm}{dp}\,W(x,\,p;\,\tau) \quad \mbox{and} \quad
\frac{1}{\rho_{\psi}(x;\,\tau)}\int_{-\infty}^{+\infty}\hspace{-.2 cm}{dp}\,J_x(x,\,p;\,\tau),
\label{zeqnz54}
\end{equation}
and even the pressure-like density can be identified by the averaged square deviation of the momentum as
\begin{equation}
\bigg{(}\int_{-\infty}^{+\infty}\hspace{-.2 cm}{dp}\,p\,J_x(x,\,p;\,\tau)\bigg{)} - \rho_{\psi}(x;\,\tau)(v^{\psi}_x(x;\,\tau))^2.
\label{zeqnz56}
\end{equation}}.

\paragraph{Non-classicality --}

The analogies with fluid dynamics are indeed much more intuitive in the classical regime.
From the phase-space coordinate vector $\mbox{\boldmath $\xi$} = (x,\,p)$, one can identify the classical phase-space velocity, 
$d{\mbox{\boldmath $\xi$}}/d\tau = \mathbf{v}_{\xi} = (v_x,\,v_p)$, so to have the flow field $\mathbf{J} = \mathbf{v}_{\xi}\,W$, with $v_x = d{x}/d\tau = p/m$ and $v_p = d{p}/d\tau = -\partial U/\partial x$.

An overall distinction between locally (Liouvillian) and globally conservative systems can be evinced through the theorem for the {\em rate of change of the volume integral bounded by a comoving closed surface} (with arbitrary velocity $\mathbf{v}_{\xi}$) around the quantity $\rho$ (c.f. Eq.~(10.811) in Ref.~\cite{Gradshteyn}), which sets the relation
\begin{equation}
\frac{D~}{D\tau} \int_{V}dV\,\rho \equiv 
\int_{V}dV\,\left[\frac{D\rho}{D\tau} +  \rho \mbox{\boldmath $\nabla$}_{\xi}\cdot \mathbf{v}_{\xi}\right]\label{zeqnz57C},
\end{equation}
with $dV \equiv dx\,dp$, and where the material derivative \cite{Steuernagel3} operator is given by
\begin{equation}
\frac{D~}{D\tau} \equiv \frac{\partial~}{\partial\tau} + \mathbf{v}_{\xi}\cdot\mbox{\boldmath $\nabla$}_{\xi}.
\end{equation}

By identifying $ \mathbf{v}_{\xi}$ with the classical phase-space vector velocity, $\mathbf{v}_{\xi(\mathcal{C})} = (p/m,\, -\partial U/\partial x)$, along a two-dimensional classical path, $\mathcal{C}$, one has for the Wigner function:
\begin{equation}
\frac{D W}{D\tau} = - W\, \mbox{\boldmath $\nabla$}_{\xi} \cdot \mathbf{v}_{\xi(\mathcal{C})},
\label{zeqnz57B}
\end{equation}
and since $\mathbf{v}_{\xi(\mathcal{C})}$ for Hamiltonian systems is divergence free, i.e. $\mbox{\boldmath $\nabla$}_{\xi} \cdot \mathbf{v}_{\xi(\mathcal{C})} = 0$, the above result implies into a conservation law, ${D W}/{D\tau} = 0$, i.e. the classical fluid-analog of the flow of the Wigner function is Liouvillian and incompressible.

In order to overview the quantum distortions and compute the time evolution of the probabilities associated to the Wigner flow, one can consider two continuity equations:
\begin{eqnarray}
\frac{\partial W}{\partial \tau} + \mbox{\boldmath $\nabla$}_{\xi}\cdot(\mathbf{v}_{\xi(\mathcal{C})}\,W )&=& 0, \qquad (classical),\\
\frac{\partial W}{\partial \tau} + \mbox{\boldmath $\nabla$}_{\xi}\cdot\mathbf{J}&=& 0, \qquad (quantum),
\label{con}
\end{eqnarray}
where the quantum current, $\mathbf{J} = \mathbf{w}\,W$, corresponds to a non-Liouvillian flow, $\mbox{\boldmath $\nabla$}_{\xi} \cdot \mathbf{w} \neq 0$.
The Wigner phase-velocity, $\mathbf{w}$, is the quantum analog of $\mathbf{v}_{\xi(\mathcal{C})}$, and exhibits a subtle unbounded divergent behavior,
\begin{equation}
\mbox{\boldmath $\nabla$}_{\xi} \cdot \mathbf{w} = \frac{W\, \mbox{\boldmath $\nabla$}_{\xi}\cdot \mathbf{J} - \mathbf{J}\cdot\mbox{\boldmath $\nabla$}_{\xi}W}{W^2},
\label{zeqnz59}
\end{equation}
given that $\mbox{\boldmath $\nabla$}_{\xi}\cdot\mathbf{J} = W\,\mbox{\boldmath $\nabla$}_{\xi}\cdot\mathbf{w}+ \mathbf{w}\cdot \mbox{\boldmath $\nabla$}_{\xi}W$ \cite{Liouvillian}.
Conditions that circumstantially imply that $\mbox{\boldmath $\nabla$}_{\xi} \cdot \mathbf{w} = 0$ are very helpful in identifying approximately Liouvillian-like trajectories in the phase-space \cite{Steuernagel3}. 

More relevantly, aiming to quantify the departure from the classical behavior of Wigner flows, for periodic motions, one can attribute a parameterization to the two-dimensional volume boundary in terms of the classical path, $\mathcal{C}$.
In this instance, the integration of the time change of $W$, $\partial W/\partial\tau$, over a volume $V_{\mathcal{C}}$ enclosed by such an oriented path, $\mathcal{C}$, yields
\begin{equation}
\int_{V_{_{\mathcal{C}}}}dV\, \frac{\partial W}{\partial \tau} =
\int_{V_{_{\mathcal{C}}}}dV\,  \left(\frac{DW}{D\tau} - \mathbf{v}_{\xi(\mathcal{C})} \cdot \mbox{\boldmath $\nabla$}_{\xi} W\right) =
\frac{D~}{D\tau}\int_{V_{_{\mathcal{C}}}}dV \,W  - \int_{V_{_{\mathcal{C}}}}dV \,\mbox{\boldmath $\nabla$}_{\xi}\cdot (\mathbf{v}_{\xi(\mathcal{C})}W), 
\label{zeqnz51BB}
\end{equation}
which can be substituted by the integral version of Eq.~(\ref{zeqnz51}) as given by the variation of the integrated probability flux, $\sigma_{(\mathcal{C})}$, enclosed by $\mathcal{C}$,
\begin{equation}
 \frac{D~}{D\tau}\sigma_{(\mathcal{C})} =\frac{D~}{D\tau}\int_{V_{_{\mathcal{C}}}}dV \,W =  \int_{V_{_{\mathcal{C}}}}dV \,\left[\mbox{\boldmath $\nabla$}_{\xi}\cdot (\mathbf{v}_{\xi(\mathcal{C})}W) - \mbox{\boldmath $\nabla$}_{\xi}\cdot \mathbf{J}\right],
\label{zeqnz51CC}
\end{equation}
which vanishes when $\mathbf{J}$ is identified to $\mathbf{v}_{\xi(\mathcal{C})}W$, i.e. in the classical limit.
Of course, the conservation of probabilities sets vanishing values for ${D}\sigma_{(\mathcal{C})}/D\tau$ in the limit of $V_{_{\mathcal{C}}} \to \infty$. 
Otherwise, by identifying the quantum corrections in terms of $\Delta \mathbf{J} = \mathbf{J} - \mathbf{v}_{\xi(\mathcal{C})}W$, one can compute the volume variation of $\sigma_{(\mathcal{C})}$ in terms of a path integral given by
\begin{equation}
 \frac{D~}{D\tau}\sigma_{(\mathcal{C})} = -\int_{V_{_{\mathcal{C}}}}dV\,  \mbox{\boldmath $\nabla$}_{\xi}\cdot \Delta \mathbf{J} = -\oint_{\mathcal{C}}d\ell\, \Delta\mathbf{J}\cdot \mathbf{n},
\label{zeqnz51DD}
\end{equation}
where the unitary vector $\mathbf{n}$ is defined by $\mathbf{n}= (-d{p}_{_{\mathcal{C}}}/d\tau, d{x}_{_{\mathcal{C}}}/d\tau) \vert\mathbf{v}_{\xi(\mathcal{C})}\vert^{-1}$, in order to set $\mathbf{n}\cdot\mathbf{v}_{\xi(\mathcal{C})}=0$.
By parameterizing the line element, $d\ell$, as $d\ell \equiv \vert\mathbf{v}_{\xi(\mathcal{C})}\vert d\tau$, one has
\begin{equation}
\frac{D~}{D\tau}\sigma_{(\mathcal{C})}
\bigg{\vert}_{\tau = T} = -\oint_{\mathcal{C}}d\ell\, \Delta\mathbf{J}\cdot \mathbf{n} = -
\int_{0}^{T}d\tau\, \Delta J_p(x_{_{\mathcal{C}}}\bb{\tau},\,p_{_{\mathcal{C}}}\bb{\tau};\tau)\,\,\frac{d}{d\tau}{x}_{_{\mathcal{C}}}\bb{\tau},
\label{zeqnz51EE}
\end{equation}
where $x_{_{\mathcal{C}}}\bb{\tau}$ and $p_{_{\mathcal{C}}}\bb{\tau}$ are typical classical solutions, $d{x}_{_{\mathcal{C}}}/d\tau = p_{_{\mathcal{C}}}\bb{\tau}/m$, $T$ is the period of the classical motion, and $\Delta J_p(x,\,p;\tau)$ is
given by Eq.~\eqref{zeqnz500}.
For stationary states, one has $\partial W/\partial \tau = 0$, and Eq.~\eqref{zeqnz51EE} describes the way that classical paths are deformed by quantum effects that are introduced by local fluid perturbations associated to flow stagnation points \cite{Steuernagel3}.

\paragraph{Entropy flux --}

The operator $\mbox{\boldmath $\nabla$}_{\xi} \cdot \mathbf{w}$ has also a closed relationship with the averaged value of the rate of change of the von Neumann entropy \cite{Entro02},
\begin{equation}
{S}_{vN} =-\int_{V}dV\, W\,\ln(W),
\label{zeqnz60}
\end{equation}
which admits a straightforward interpretation for positive definite Wigner functions.
From Eq.~(\ref{zeqnz57C}), one obtains
\begin{eqnarray}
\frac{D{S}_{vN}}{D\tau} &=& -\frac{D~}{D\tau}\left(\int_{V}dV\, W\,\ln(W)\right)\nonumber\\
&=& -\int_{V}dV\,\left[\frac{D~}{D\tau} (W\,\ln(W)) + W\,\ln(W) \mbox{\boldmath $\nabla$}_{\xi}\cdot \mathbf{w}\right]\nonumber\\
&=&- \int_{V}dV\,\left[\frac{\partial~}{\partial \tau} (W\,\ln(W)) + \mbox{\boldmath $\nabla$}_{\xi}\cdot(\mathbf{w} W\,\ln(W))\right],
\label{zeqnz61}
\end{eqnarray}
which, after rewriting $\partial W/\partial \tau$ in terms of $\mathbf{J} = \mathbf{w}\,W$ (c.f. Eq.~(\ref{con})), results into
\begin{eqnarray}
\frac{D{S}_{vN}}{D\tau} &=& \int_{V}dV\,\left[\mbox{\boldmath $\nabla$}_{\xi}\cdot\mathbf{J} + \ln(W)\,\mbox{\boldmath $\nabla$}_{\xi}\cdot\mathbf{J}-\mbox{\boldmath $\nabla$}_{\xi}\cdot(\mathbf{J}\,\ln(W))\right]
\nonumber\\
&=& \int_{V}dV\,\left[\mbox{\boldmath $\nabla$}_{\xi}\cdot\mathbf{J} -W^{-1}\mathbf{J}\cdot\mbox{\boldmath $\nabla$}_{\xi}W\right]
= \int_{V}dV\,\left[W \,\mbox{\boldmath $\nabla$}_{\xi}\cdot\mathbf{w}\right]
\equiv \langle \mbox{\boldmath $\nabla$}_{\xi}\cdot\mathbf{w} \rangle,
\label{zeqnz62}
\end{eqnarray}
where $\langle \mbox{\boldmath $\nabla$}_{\xi}\cdot\mathbf{w} \rangle = Tr_{\{x,p\}}\left[\hat{\rho}\mbox{\boldmath $\nabla$}_{\xi}\cdot\mathbf{w}\right]$ (c.f. Eq~\eqref{five}).
This relationship can be cast into the form of a continuity equation,
\begin{equation}
\frac{D{S}_{vN}}{D\tau} - \langle \mbox{\boldmath $\nabla$}_{\xi}\cdot\mathbf{w} \rangle =0,
\label{zeqnz62BB}
\end{equation}
which sets the global entropy dynamics in the limit of $V \to \infty$.
Again, one should notice that $\partial (W\,\ln(W))/\partial \tau = 0$ for stationary states, 
and then, from Eq.~\eqref{zeqnz61}, once $V$ is identified with $V_{_{\mathcal{C}}}$, one has
\begin{eqnarray}
\frac{D~}{D\tau}{S}_{vN(\mathcal{C})}
\bigg{\vert}_{\tau = T} &=& \oint_{\mathcal{C}}d\ell\, \ln(W)\left(\mathbf{J}\cdot \mathbf{n}\right) \nonumber\\
&=&
\int_{0}^{T}d\tau\, \ln(W(x_{_{\mathcal{C}}}\bb{\tau},\,p_{_{\mathcal{C}}}\bb{\tau};\tau))\,\,\Delta J_p(x_{_{\mathcal{C}}}\bb{\tau},\,p_{_{\mathcal{C}}}\bb{\tau};\tau)\,\,\frac{d~}{d\tau}{x}_{_{\mathcal{C}}}\bb{\tau},
\label{zeqnz62CC}
\end{eqnarray}
for a volume $V_{\mathcal{C}}$ enclosed by $\mathcal{C}$, and for the dynamics driven by conservative systems (i.e. Hamiltonians with a potential $U\equiv U(x)$).

\paragraph{Purity flux --}

A first approach for computing the quantum entropy content of the Wigner function can be interestingly achieved introducing an additional contribution given by $-\ln(2\pi)$, that is:
\begin{eqnarray}
-\ln(2\pi) - \int_{V}dV\, W\,\ln(W) &=& -\int_{V}dV\, W\,\ln(2\pi \,W) 
= \int_{V}dV\, W - 2\pi \int_{V}dV\, W^2 + \dots\nonumber\\ &=& 1 - \mathcal{P} +\dots,
\label{zeqnz63}
\end{eqnarray}
from which is possible to identify a rate of change of purity, $\mathcal{P}$, as given by
\begin{eqnarray}
\frac{1}{2\pi}\frac{D\mathcal{P}}{D\tau} &=& \frac{D~}{D\tau}\left(\int_{V}dV\, W^2\right)= \int_{V}dV\,\left[\frac{D~}{D\tau} W^2 + W^2 \mbox{\boldmath $\nabla$}_{\xi}\cdot \mathbf{w}\right]\nonumber\\
&=& \int_{V}dV\,\left[\frac{\partial~}{\partial \tau}W^2 + \mbox{\boldmath $\nabla$}_{\xi}\cdot(\mathbf{w} W^2)\right]=\int_{V}dV\,\left[\mbox{\boldmath $\nabla$}_{\xi}\cdot(W\,\mathbf{J}) - 2 W\,\mbox{\boldmath $\nabla$}_{\xi}\cdot\mathbf{J}\right]\nonumber\\
&=& -\int_{V}dV\,\left[W\,\mbox{\boldmath $\nabla$}_{\xi}\cdot\mathbf{J} - \mathbf{J}\cdot\mbox{\boldmath $\nabla$}_{\xi}W\right]
= -\int_{V}dV\,\left[W^2 \,\mbox{\boldmath $\nabla$}_{\xi}\cdot\mathbf{w}\right]
\nonumber\\
&\equiv& - \langle W\,\mbox{\boldmath $\nabla$}_{\xi}\cdot\mathbf{w} \rangle,
\label{zeqnz64}
\end{eqnarray}
which also can be cast in the form of
\begin{equation}
\frac{1}{2\pi}\frac{D\mathcal{P}}{D\tau} + \langle W\,\mbox{\boldmath $\nabla$}_{\xi}\cdot\mathbf{w} \rangle =0.
\label{zeqnz64BB}
\end{equation}
In this case, $\partial (W^2)/\partial \tau = 0$ for stationary states and one has 
\begin{eqnarray}
\frac{D~}{D\tau}\mathcal{P}_{(\mathcal{C})}
\bigg{\vert}_{\tau = T} &=& -\oint_{\mathcal{C}}d\ell\, W\,\mathbf{J}\cdot \mathbf{n} \nonumber\\&=& -
\int_{0}^{T}d\tau\, W(x_{_{\mathcal{C}}}\bb{\tau},\,p_{_{\mathcal{C}}}\bb{\tau};\tau) \,\Delta J_p(x_{_{\mathcal{C}}}\bb{\tau},\,p_{_{\mathcal{C}}}\bb{\tau};\tau)\,\,\frac{d~}{d\tau}{x}_{_{\mathcal{C}}}\bb{\tau},
\label{zeqnz64DD}
\end{eqnarray}
for the boundary $\mathcal{C}$ encompassing the classical trajectories.

Loss and production rates of $S_{vN}$ and $\mathcal{P}$ are driven by quantum distortions over a Liouvillian background flow.
Furthermore, from Eq.~(\ref{zeqnz500}), it is possible to demonstrate that
\begin{eqnarray}
\frac{D\mathcal{P}}{D\tau} \propto \int_{-\infty}^{+\infty}\hspace{-.3 cm}dp\, W\,\left(\frac{\partial~}{\partial p}\right)^{2k+1}\hspace{-.5 cm} W(x,\,p;\,\tau) = 0,
\label{zeqnz65}
\end{eqnarray}
i.e. the purity is a constant of the motion when the integration over the entire volume is extended to $\infty$ or over a symmetric interval in the momentum direction, for the cases where $W$ is symmetric in $p$.
Thus, purity can only be locally affected.

To sum up, all the above tools measure how quantum and classical systems differ from each other, or more properly, how far the quantum regime is from the classical one. They correspond to quantifiers of several types of non-classicality, each of them related to probability, entropy and purity fluxes, through the respective continuity equation.
For the quantum harmonic oscillator Liouvillian system, all the effects vanish since the corresponding integrals are reduced to path integrals of $d{x}_{_{\mathcal{C}}}\bb{\tau}/d\tau$ (multiplied by an arbitrary constant), which is null for an enclosing path along the periodic motion.

\begin{figure}
\vspace{1. cm}
\includegraphics[scale=0.56]{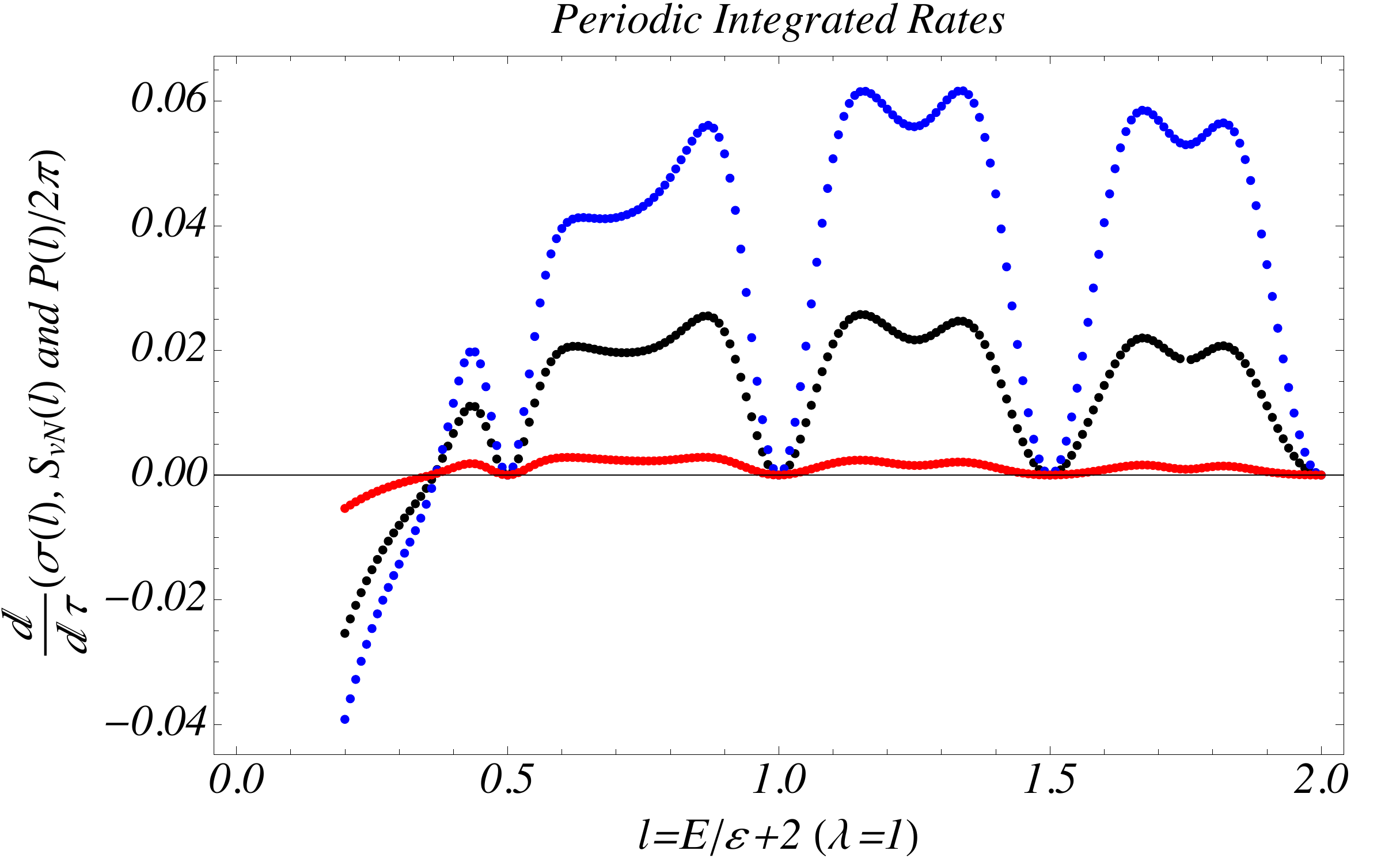}

\vspace{1. cm}
\includegraphics[scale=0.56]{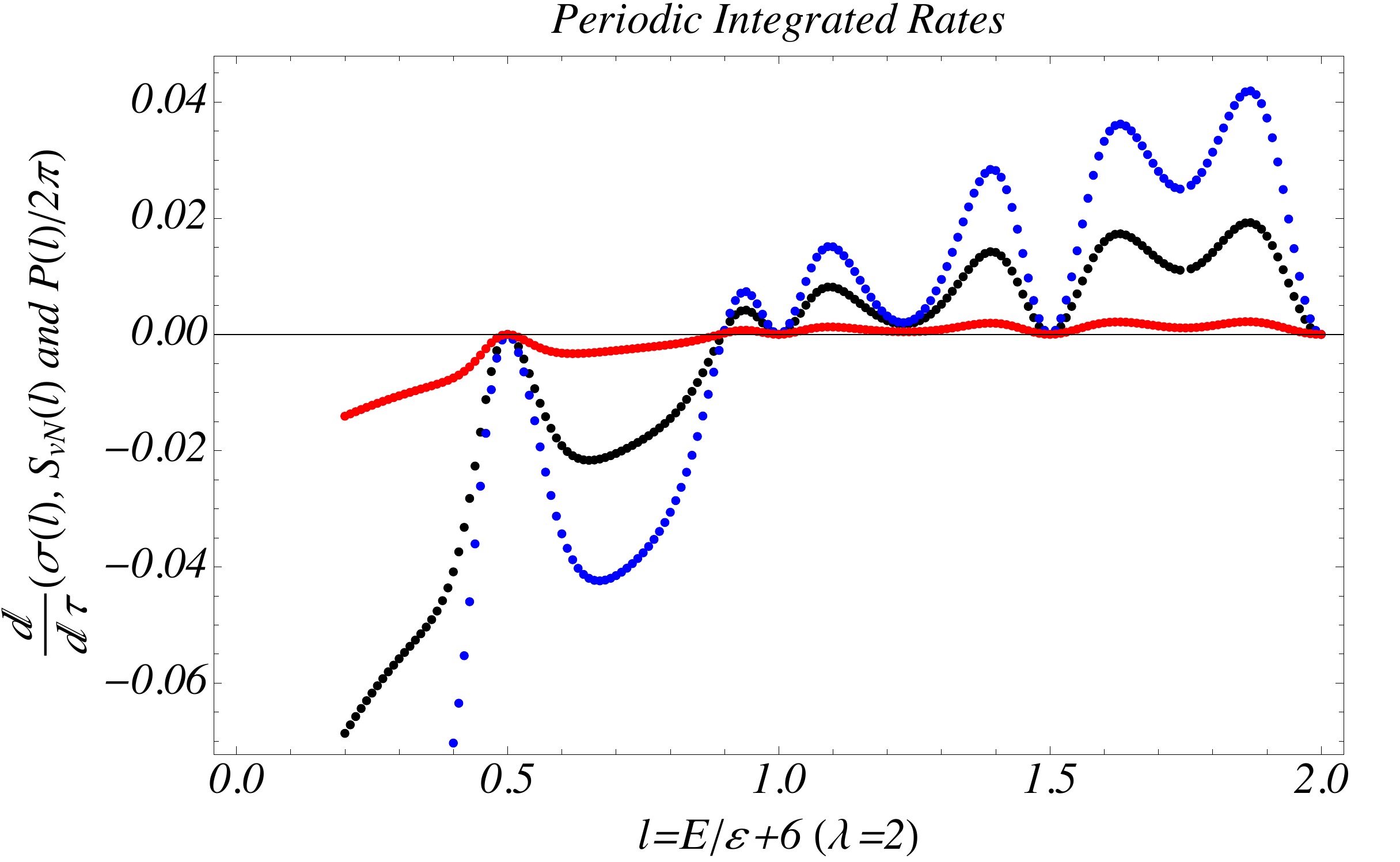}
\renewcommand{\baselinestretch}{.85}
\caption{
(Color online) Quantifiers of decoherence (black), entropy flux (blue) and purity flux (red) for the periodic anharmonic system driven by a hyperbolic PT potential (c.f. Eq.~\eqref{Ham001}) as function of the total energy parameter $l = E/\varepsilon + \lambda(\lambda+1)$ for quantum numbers $\lambda = 1$ (first plot) and $\lambda =2$ (second plot).
The results correspond to the rates of local transference (throughout the boundary surface encompassing the classical path, $\mathcal{C}$) of information, respectively expressed by $\frac{D~}{D\tau}\sigma_{(\mathcal{C})}
{\vert}_{\tau = T} $, $\frac{D~}{D\tau}{S}_{vN(\mathcal{C})}
{\vert}_{\tau = T}$, and $\frac{D~}{D\tau}\mathcal{P}_{(\mathcal{C})}
{\vert}_{\tau = T}$, in a time interval corresponding to the classical period of motion, $T = 2\pi/{l}$.}
\label{Figura001}
\end{figure}

In order to quantify the quantum distortions onto a classical background, a typical quantum system for which the above quantifiers can be computed is, for instance, the one described by the hyperbolic P\"oschl-Teller (PT) anharmonic solutions \cite{PT,Curtright,Bund}, where the Hamiltonian is given by
\begin{equation}
H = \frac{p^{2}}{2m} - \varepsilon \,\lambda(\lambda + 1) \mbox{sech}^2(x/L),
\label{Ham001}
\end{equation}
with $\varepsilon = \hbar^2/2mL$.
The canonical variables can be rewritten in terms of dimensionless quantities $s \equiv x/L$ and $q\equiv p L/\hbar$, as to have $[s,\,q]= i$ and a simpler dimensionless version of $H$ given by
\begin{equation}
H_{\varepsilon} = H/\varepsilon =  q^2 - \lambda(\lambda + 1) \,\mbox{sech}^2(s),
\label{Ham002}
\end{equation}
such that periodic classical solutions, for $H < 0$, with $s(0) =0$ and $q(0)= \sqrt{l}$, with arbitrary $l$, are given, in terms of a dimensionless time, $\tau = 2\varepsilon t/\hbar$, by
\begin{eqnarray}
\label{clas1}
s(\tau) &=& \pm  \mbox{arcsinh}\left[(1/\sqrt{l})\, \sin(l\tau)\right],\\
\label{clas2}
q(\tau) &=& \pm  \frac{l\,\cos(l \tau)}{\sqrt{l+ \sin^2(l \tau)}},
\end{eqnarray}
and, for instance, the ground-state (positive definite) Wigner function is given by \cite{Curtright}
\begin{eqnarray}
\label{p1wBBBB}
W_{\lambda}(s,\,q) &=&
\frac{2}{\sqrt{\pi}}\frac{\Gamma(\lambda+\frac{1}{2})}{\Gamma^2(\lambda)}
\mathcal{D}^{\lambda -1}_{(s)}\,f(s,\,q),
\end{eqnarray}
with
\begin{equation}
\mathcal{D}_{(s)} \equiv \frac{(-1)}{\sinh(2s)}\frac{d\,}{ds},
\qquad \mbox{and} \qquad
f(s,\,q) 
= \frac{\sin(2\,q \,s)}{\sinh(2s)\,\sinh(\pi q)}.
\end{equation}
Applying the definition Eq.~\eqref{zeqnz500} onto the above quantum system in the ground state configuration and using Eqs.~\eqref{zeqnz51EE}, \eqref{zeqnz62CC} and \eqref{zeqnz64DD}, one gets the integrated results for {\em probability, entropy and purity fluxes} according to the Wigner flow framework as depicted in Fig.~\ref{Figura001}.
Notice that the results are quantitatively consistent with each other. These integrated quantifiers of non-classicality all depict that the classical to quantum discrepancies increase with the associated energy parameter, $l = E/\varepsilon + \lambda(\lambda+1)$, reaching a maximal value and decreasing. 
Interestingly, the {\em classical versus quantum discrepancy effects} are suppressed for quantized values of $E = (n/2-\lambda(\lambda+1))\varepsilon  < 0$, with $n$ and $\lambda$ assuming integer values. 
For instance, in the case where $l=\lambda$, one has $E= -\lambda^2\varepsilon$, which is the energy for the ground state of the quantum problem associated to $\lambda$, and the classical trajectories are identified by an energy equals to $-\lambda^2\varepsilon$. 
Classical trajectories in the phase-space only accommodate -- without yielding quantum discrepancies -- the quantized energy version of the quantum system.
Once enclosed by classical trajectories with energies matching the quantized ones, the measure of non-classicality of the quantum system is expressed by the nodes in the plots from Fig.~\ref{Figura001}.
It means that the finite volume of the phase-space, $V_\mathcal{C}$, which is enclosed by $\mathcal{C}$, comprise quantum stagnation points which sum a vanishing gross effect.
The nodes are indicative of quantization and not of non-classicality.

Of course, this is only an example of such a tool which can be applied to the emergence of quantum states and to qualitatively express quantum to classical transitions.
Although specialized for the example of the PT Hamiltonian \cite{PT,Curtright}, our results are universal and useful as quantifiers of quantum decoherence for any quantum state described by a continuous Wigner function. It is much more general than quantifiers involving gaussian covariance approaches.

To conclude, our results show that the quantum information profile of quantum systems, once driven by  decoherence, von Neumann entropy and purity quantifiers, can be cast into the form of continuity equations in the context of the WW formalism of QM in the phase-space.
Furthermore, local aspects of the Wigner flow have been computed as to quantitatively express the phase-space features and the non-Liouvillian nature of quantum systems.
Our results are applicable to any quantum system which admits a description in terms of the WW formalism. {For instance, it has been extended to the evaluation of modified Laguerre Wigner functions in a procedure which describes the classical to quantum transition in the context of quantum cosmology} \cite{NossoPaper}, {and it shall also be considered in the forthcoming investigation of Wigner functions for an electron in the Coulomb potential.}

{\em Acknowledgments: The work of AEB is supported by the Brazilian Agency FAPESP (grant 17/02294-2).}

\end{document}